# Design and Implementation of IEEE 802.15.4 Mac Protocol on FPGA

Naagesh S. Bhat
Student
M.S.Ramaiah School of Advanced Studies

## ABSTRACT

The IEEE 802.15.4 is a wireless standard introduced for low power, low cost wireless communication with moderate data rates. In the next few years, it is expected that Low Rate Wireless Personal Area Networks (LR-WPAN) will be used in a wide variety of embedded applications, including home automation, industrial sensing and control, environmental monitoring and sensing. In these applications, numerous embedded devices running on batteries are distributed in an area communicating via wireless radios. This work presents a method which can be used for comparing current consumption of wireless data transfer embedded systems. This paper implements a small subset of the IEEE 802.15.4 protocol to achieve a point to point communication. The implemented protocol uses 802.15.4 MAC compliant data and acknowledgment packets. Current consumption is measured while doing one data packet transmission. Measurements are compared with existing work. IEEE 802.15.4 protocol implementation is done using Verilog language. Code implementation is done in such a manner so that it can be ported to any platform with minimal changes. It can also be modified to suit any special experimental setup requirements.

**Keywords**: IEEE 802.15.4, Low Rate – Wireless Personal Area Network, Point to Point Communication

## 1. INTRODUCTION

IEEE Standard 802.15.4[1] defines the physical layer (**PHY**) and medium access control (**MAC**) sub layer specifications for low-data-rate wireless connectivity with fixed, portable, and moving devices with no battery or very limited battery consumption. Wireless Personal Area Networks (**WPAN**s) are used to convey information over relatively short distances. The infrastructure is very less compared to Wireless Local Area Network. So a Low Rate WPAN (**LR-WPAN**) is a simple, low-cost communication network that allows wireless connectivity in applications with limited power and relaxed throughput requirements. Ease of installation, reliable data transfer, short-range operation, extremely low cost, and a reasonable battery life, while maintaining a simple and flexible protocol are the main objectives of LR-WPAN. This project work is to implement IEEE 802.15.4 MAC Protocol on to FPGA. It also includes the interaction of the blocks with Zigbee transceivers which can transmit or receive the data. Full-function device (**FFD**) and a reduced-function device (**RFD**) are the two different device types that can participate in an IEEE 802.15.4 network.

In Section 2, discusses about the background theory for IEEE 802.15.4 and in specific to PHY layer and MAC Sub-Layer Specifications. The design of 802.15.4 protocol is described in Section 4. Results of IEEE 802.15.4 protocol are discussed in Section 5.

## 2. BACKGROUND

IEEE 802.15.4 is a member of the IEEE 802 family, but it does not mean that all the features of all the other IEEE 802 standards are included or even desired for this low-rate, low-duty cycle standard. Control of expectations is probably one of the greatest challenges for any standards development organization, and this standard is no exception. The mission for this standard was to empower simple devices with a reliable, robust wireless technology that could run for years on standard primary batteries, was designed to allow a developer who had little ability or interest in the radio technology or communications protocol arts to effectively use and benefit from radios based upon the standard. IEEE 802.15.4 protocol stack is shown in Fig. 1

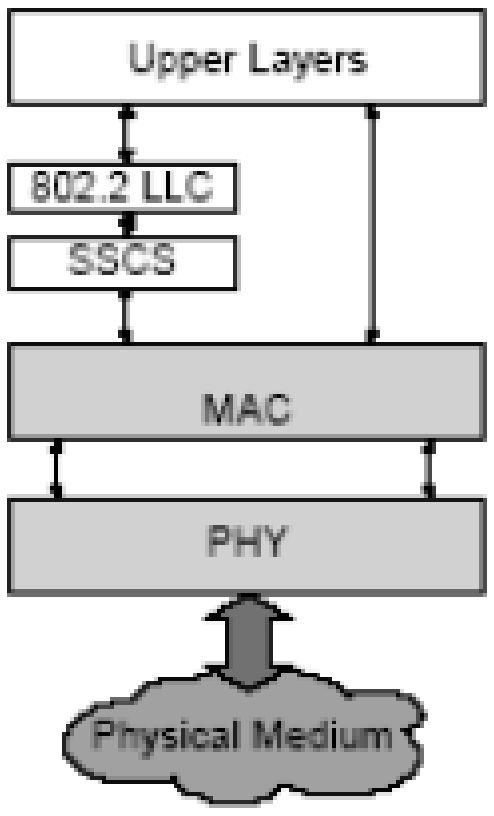

**Fig 1: IEEE 802.15.4 Protocol Stack**

By favoring low-cost and low-power, IEEE 802.15.4 is enabling applications in the fields of industrial, agricultural, vehicular, residential and medical sensors and actuators. Until recently, these applications could not make use of current wireless technologies or would have to use proprietary solutions (in most cases unidirectional) [2,3]. The intent of IEEE 802.15.4 is to address applications where existing WPAN solutions are too expensive and the performance of a technology such as Bluetooth is not required. IEEE 802.15.4 LR-WPANs complement other WPAN technologies by providing very low power consumption capabilities at very low cost, thus enabling applications that were previously impractical. Table 1 illustrates a basic comparison between IEEE 802.15.4 and other IEEE 802 wireless networking standards.

**Table 1. IEEE 802 Comparison**

| | 802.11b WLAN | 802.15.1 WPAN | 802.15.4 LR-WPAN |
|---|---|---|---|
| Range | ~100 m | ~10 - 100 m | 10 m |
| Raw Data Rate | 11 Mbps | 1 Mbps | <= 0.25 Mbps |
| Power Consumption | medium | low | ultra low |

A summary of the high-level features of the IEEE 802.15.4 is shown in Table 2.

**Table 2. IEEE 802.15.4 High Level Characteristics**

| | | | | |
|---|---|---|---|---|
| Frequency Band | Two PHYs | Low-Band (BPSK) | 868 MHz | 1 channel - 20 kb/s |
| | | | 915 MHz | 10 channels - 40 kb/s |
| | | High-Band (O-QPSK) | 2.4 GHz | 16 channels - 250 kb/s |
| Channel Access | CSMA-CA and slotted CSMA-CA | | | |
| Range | 10 to 20m | | | |
| Addressing | Short 8 bit or 64-bit IEEE | | | |

## 2.1.LR-WPAN Design

A main design consideration for LR-WPANs is low power consumption, thereby maximizing battery life. To achieve low average power consumption, IEEE 802.15.4 assumes that the amount of data transmitted is short and that it is transmitted infrequently in order to keep a low duty-cycle. In addition, the packet structure was designed to add minimal overhead over the transported payload. The standard allows the formation of two possible network topologies: the star topology or the peer-to-peer topology, Fig 2. In the star topology, the communication is performed between network devices and a single central controller, called the PAN coordinator. A network device is either the initiation point or the termination point for network communications. The PAN coordinator is in charge of managing all the star PAN functionality. In the peer-to-peer topology, every network device can communicate with any other within its range.

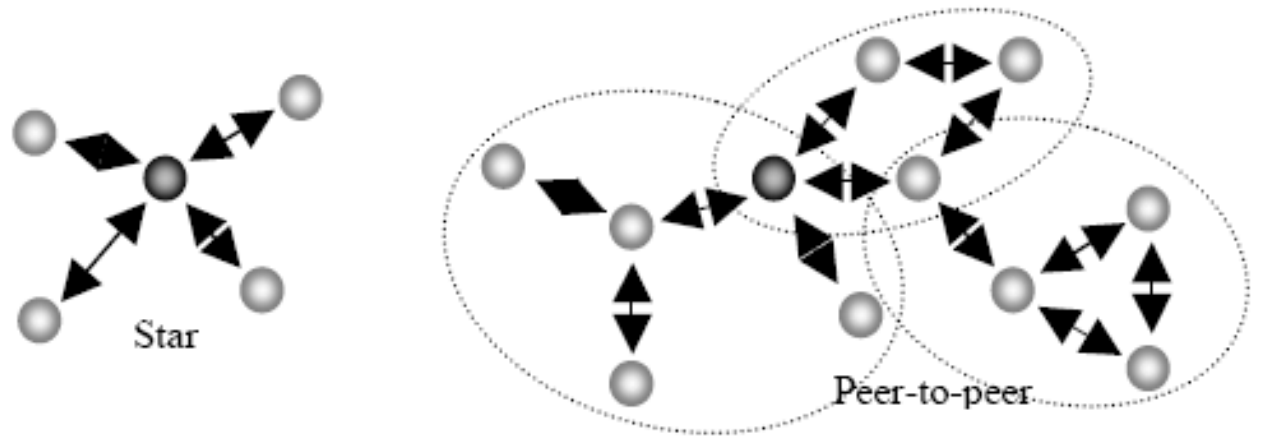

**Fig 2: Star and Peer-to-peer topology**

This topology also contains a PAN coordinator, which acts as the root of the network. Peer-to-peer topology allows more complex network formations to be implemented; e.g. ad hoc and self-configuring networks. The routing mechanisms required for multi-hopping are part of the network layer and are therefore, not in the scope of IEEE 802.15.4. An IEEE 802.15.4 LR-WPAN device is composed of a physical (PHY) layer and a medium access control (MAC) sublayer that provides access to the physical channel for all types of transfer and ensures the reliable transfer of frames.

## 2.2.PHY Layer

IEEE 802.15.4 was designed to support two PHY options based on Direct Sequence Spread Spectrum (DSSS); this characteristic allows the use of low-cost digital IC realizations. Both PHYs make use of the same basic packet structure for low-duty-cycle low-power operation. The primary difference between the two PHYs is the frequency band. The 868/915 MHz PHY (also called low-band) is specified for operation in the 868 MHz band in Europe offering one channel with a raw data rate of 20 kb/s and the 915 MHz ISM band in North America offering 10 channels with a raw data rate of 40 kb/s. The low-band uses binary phase shift key (BPSK) modulation. Both PHY layers use a common packet structure, enabling the definition of a common MAC interface. Each packet, or PHY protocol data unit (PPDU), contains a preamble, a start of packet delimiter, a packet length, and a payload field, or PHY service data unit. The 32-bit preamble is designed for acquisition of symbol and chip timing. The IEEE 802.15.4 payload length can vary from 2 to 127 bytes. This structure is shown in Fig 3.

| PHY protocol data unit (PPDU) | | | |
|---|---|---|---|
| Preamble | Start of packet delimiter | Length Field | PHY layer payload<br>PHY service data unit (PSDU) |
| 4 bytes | 1 byte | 1 byte | 2-127 bytes |

**Fig 3: IEEE 802.15.4 Packet Structure**

## 2.3.MAC Layer

The IEEE 802 project divides the data link layer (DLL) in to two sub layers, the MAC and logical link control (LLC) sub layers. The LLC is standardized in IEEE 802.2 and is common among the IEEE 802 standards. The IEEE 802.15.4 medium access control (MAC) sublayer controls the access to the radio channel employing the CSMACA mechanism. If upper layers detect that the communications throughput has been degraded below a determined threshold, the MAC will be instructed to perform an energy detection scan through the available channels. Based on the detected energy, the upper layers will switch to the channel with the lowest energy. The IEEE 802.15.4 performs the energy scan by the use of a clear channel assessment procedure. This can be performed by following either simple in-band energy detection above a threshold, or an IEEE 802.15.4 carrier detection or a combination of both. The 802.15.4 MAC is also responsible for flow control via acknowledged frame delivery, frame validations as well as maintaining network synchronization, controlling the association, administering device security and scheming the guaranteed time slot mechanism.

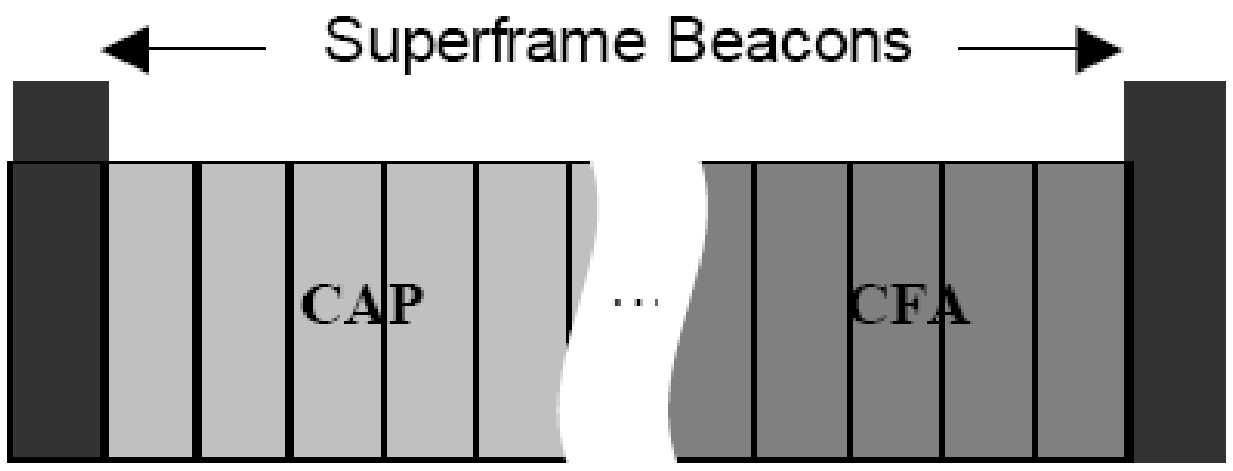

**Fig 4: Superframe Structure**

The LR-WPAN standard allows the optional use of a superframe structure for applications requiring dedicated bandwidth to guarantee communication latency. The format of the superframe is defined by the PAN coordinator, by using the network beacons which bound the superframe structure. The superframe is composed of 16 equally sized time slots grouped in two sections: the contention access period (CAP) and the contention free period (CFP). The time slots assigned for the CFP are called guaranteed time slots (GTS) and are administered by the PAN coordinator. A pictorial of the superframe structure is shown in Fig 4.

## 2.4.Frame Structure

The frame structures have been designed to keep the complexity to a minimum while at the same time making them sufficiently robust for transmission on a noisy channel. Each successive protocol layer adds to the structure with layer-specific headers and footers. This standard defines four frame structures:

- Beacon Frame, used by a coordinator to transmit beacons
- Data Frame, used for all transfers of data
- Acknowledgment Frame, used for confirming successful frame reception
- MAC Command Frame, used for handling all MAC peer entity control transfers

Depicted on Fig 5 is the PPDU frame structure. One layer above the PPDU is the physical layer. It consists of the synchronization header (SHR), the start of frame delimiter (SFD), the frame length and the MAC protocol data unit (MPDU). The SHR itself is four bytes of0x00. Next, the SFD is defined as the byte 0xA7, and the frame length is the length of the MPDU. One layer higher than the physical layer is the MAC layer, i.e., the MPDU. It consists of the frame control field (FCF), the data sequence number, the address information, the frame payload, and the frame check sequence (FCS). The explanation of the individual fields of the FCF can be found in the IEEE 802.15.4 standard. In short, the FCF tells the recipient what type of packet he just received and how the address information is stored. Depending on the type of frame, the address can be left out or consist of up to 20bytes. The FCS is the CRC-CCITT 16-bit checksum of the MPDU.

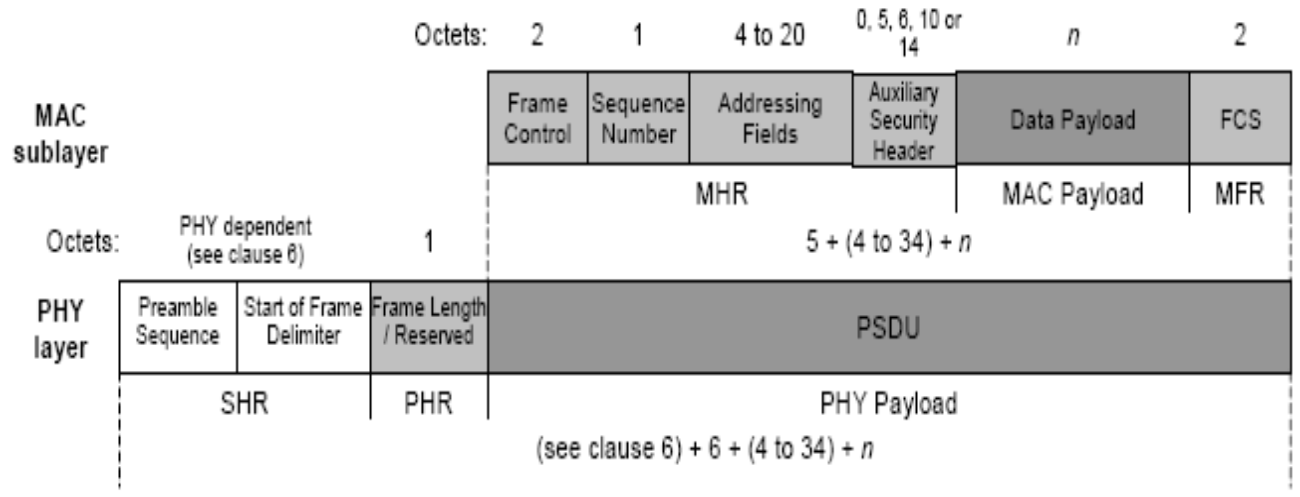

**Fig 5: Data and MAC Frame Structure**

## 3. IMPLEMENTATION

IEEE 802.15.4 works mainly on 3 different 2.4GHz, 915 MHz and 868 MHz as mentioned in Fig. 6. Out of three different modulations functions, the proposed design is implemented with 2.4GHz design.

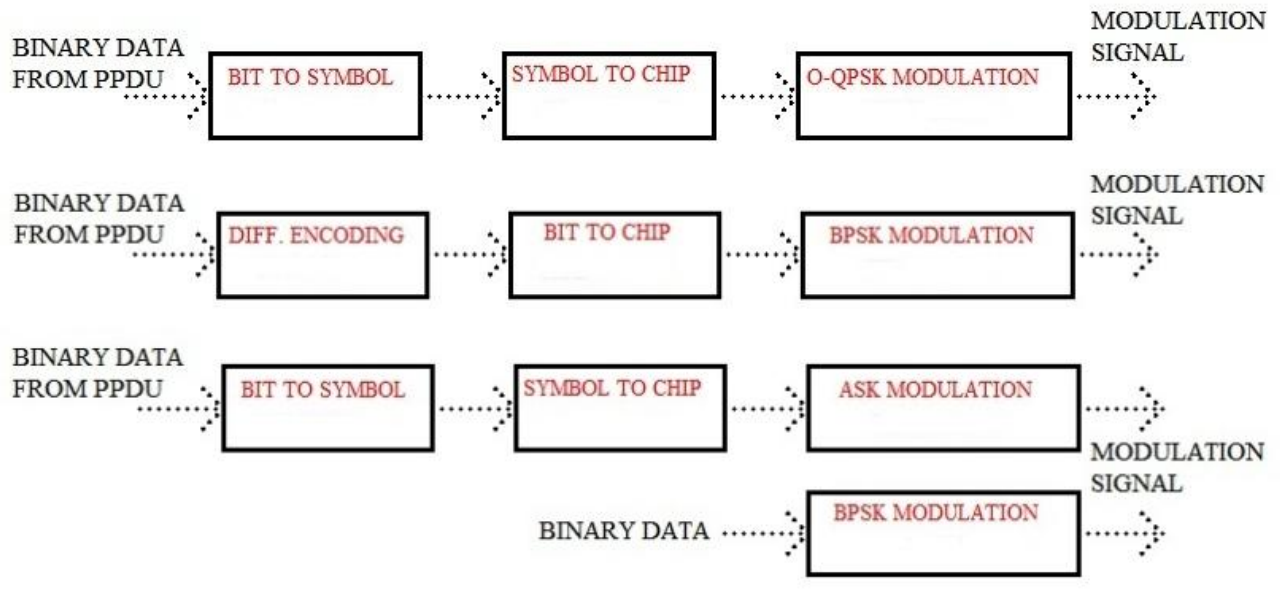

**Fig 6: IEEE 802.15.4 modulation functions as per frequencies**

## 3.1.Bit to Symbol Mapping

The 4 LSBs (b0, b1, b2, b3) of each octet shall map into one data symbol, and the 4 MSBs (b4, b5, b6, b7) of each octet shall map into the next data symbol. Each octet of the PPDU is processed through the modulation and spreading functions (see Fig. 6) sequentially, beginning with the Preamble field and ending with the last octet of the PSDU.

## 3.2.Symbol to Chip Mapping

Each data symbol shall be mapped into a 32-chip PN sequence. The PN sequences are related to each other through cyclic shifts and/or conjugation.

## 3.3.O-QPSK Modulation

The chip sequences representing each data symbol are modulated onto the carrier using O-QPSK with half sine pulse shaping. Even-indexed chips are modulated onto the in-phase (I) carrier and odd-indexed chips are modulated onto the quadrature-phase (Q) carrier. Because each data symbol is represented by a 32-chip sequence, the chip rate (nominally 2.0 Mchip/s) is 32 times the symbol rate. To form the offset between I-phase and Q-phase chip modulation, the Q-phase chips shall be delayed by *Tc* with respect to the I-phase chips (see Fig 7), where *Tc* is the inverse of the chip rate.

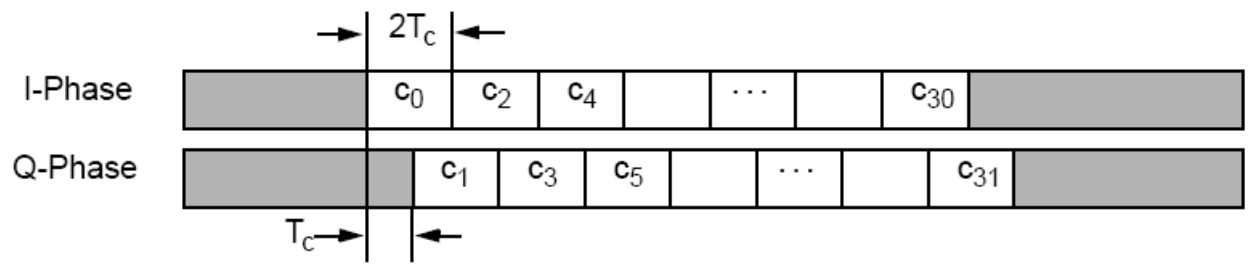

**Fig 7: Offset QPSK chip offsets**

## 3.4.Frequency Modulation

Fig 8 represents the simplified IEEE 802.15.4 transmitter section. The design depicted in this section was carried out utilizing the SINE WAVE GENERATOR to generate the SINE and COSINE waveforms. The SWG is used to generate a 16-bit SINE and COSINE values. The generated values are then added with the modulated data generated out of the In-Phase and Quadrature Phase respectively. The out signal, thus generated, is then transmitted.

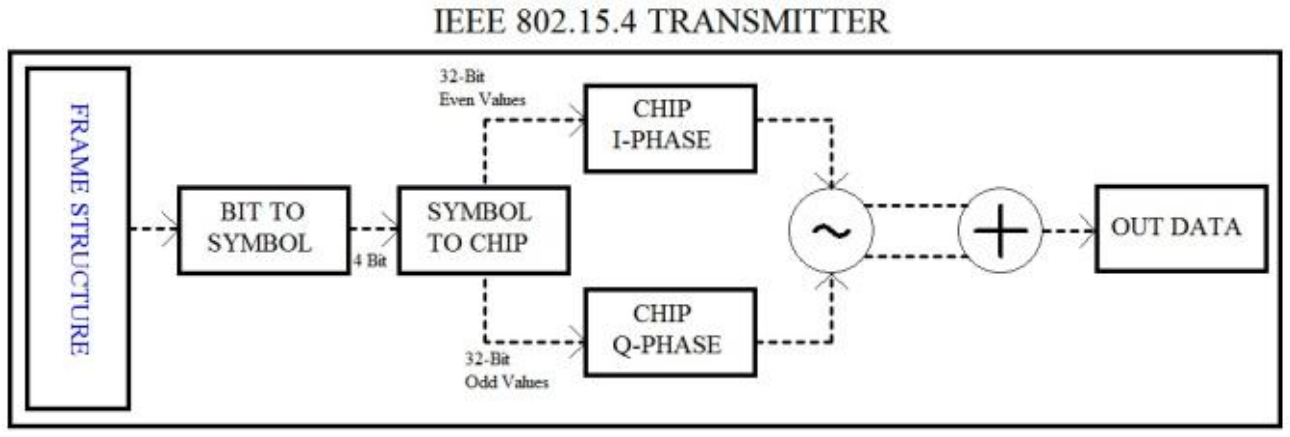

**Fig 8: IEEE 802.15.4 Transmitter Section**

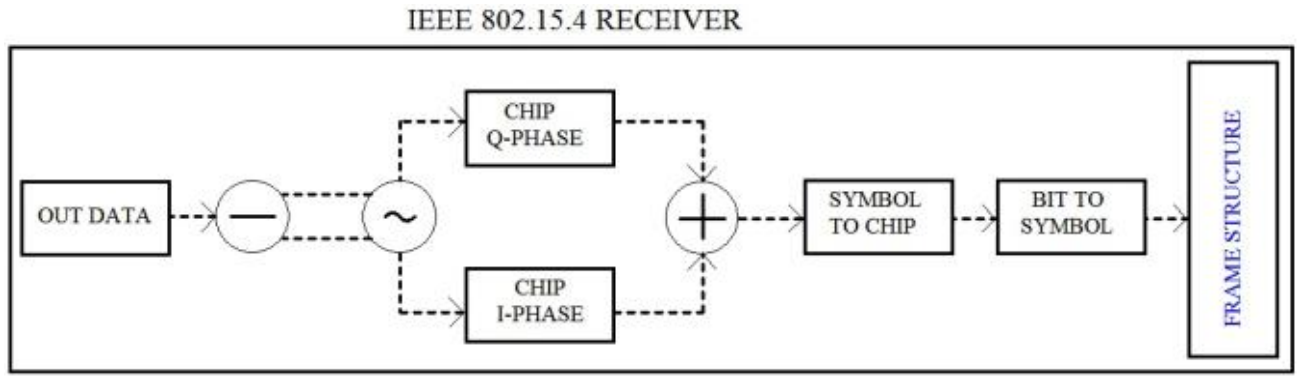

**Fig 9: IEEE 802.15.4 Receiver Section**

The receiver section, illustrated in Fig 9, then de-serializes the output signal and is passed onto the SINE WAVE generator where the SINE and COSINE values are subtracted. Once the original signal is obtained after the SINE and COSINE removal, the data is demodulated from other blocks from which the original frame structure is received. From the data structure the data is extracted after the comparison of frame control signal.

# 4. SIMULATIONS AND RESULTS

## 4.1.Test Bench Data Generation

The test-bench data for the input to the system simulation were generated and stored in a data file. The data generation procedure follows the bit-to-chip mapping rule specified by the IEEE 802.15.4 standard and the symbol-to-sample mapping rule of the offset-QPSK modulation. Each of 4-bit input data is mapped to a corresponding 32-chip sequence according to the PN-code table specified by the IEEE 802.15.4 standard. This sequence splits into two sub-sequences (channels) and is fed into the QPSK mapping module. In the QPSK mapping module, each chip is mapped into a half-sine formed by four 16-bitsamples. To produce the offset-QPSK signal, one of the two sample sequences is delayed by two sample periods. Finally, both the sample sequence is added to reduce noise.

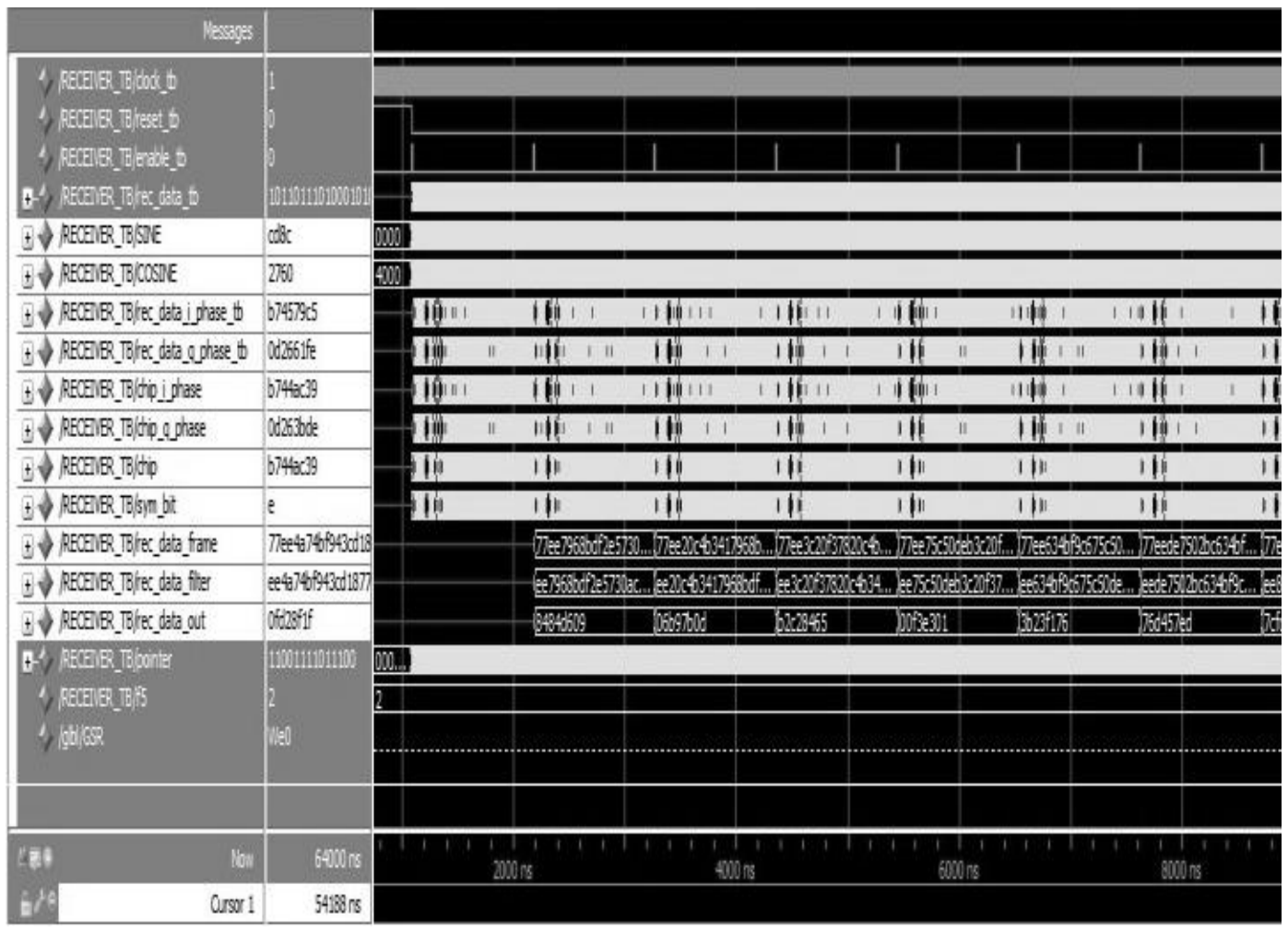

**Fig 10: Simulation Output on Receiver**

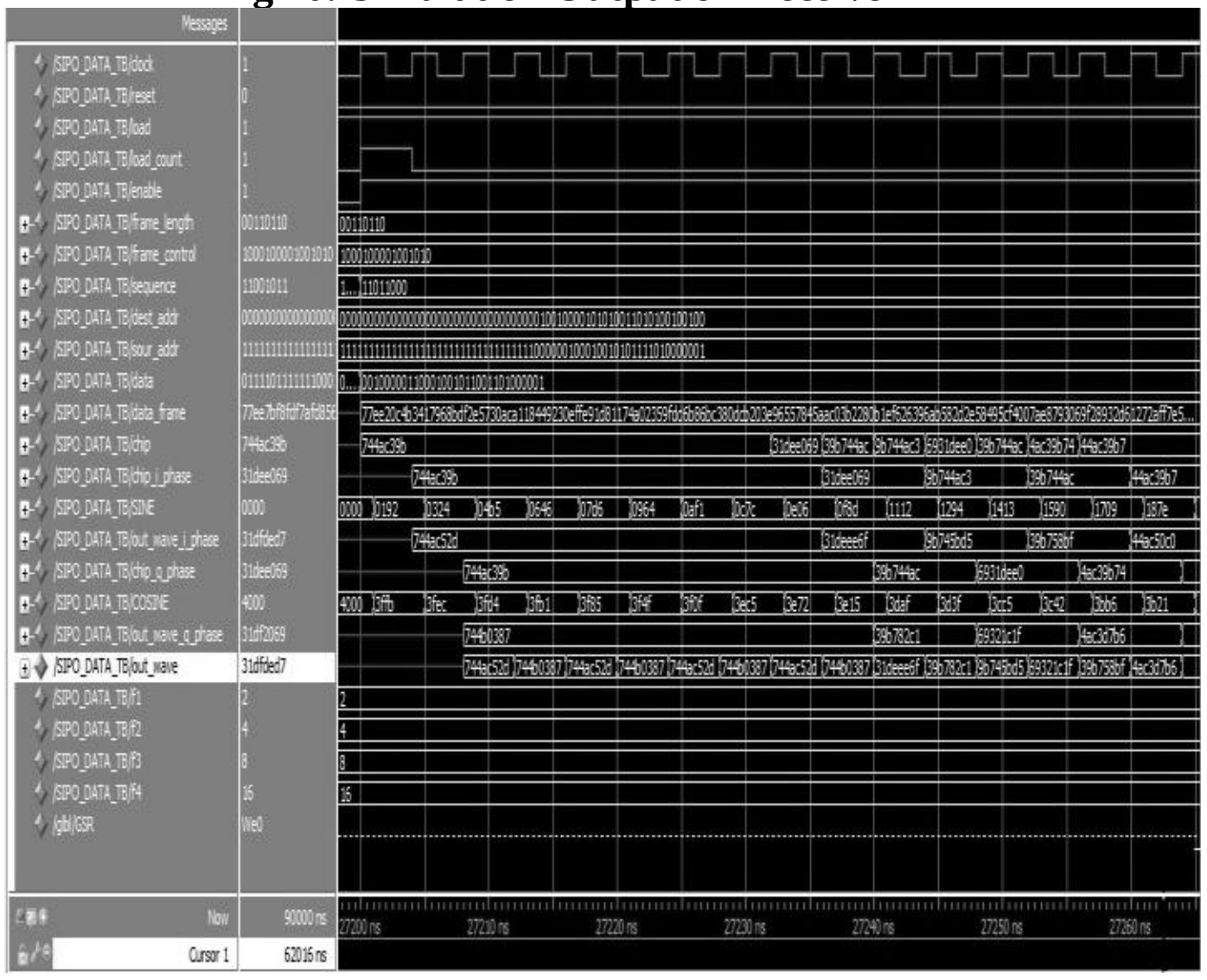

**Fig 11: Simulation Output on Transmitter**

## 4.2.Performance on Hardware

During the hardware testing of the design, some of the issues were discovered. The main issue is with respect to clock. The design is implemented in Virtex 4 Pro where the clock speed is around 100MHz whereas the design has to run only in 250 KHz. So the clock divider has to be implemented first in the hardware and then the main design.

# 5. CONCLUSION

This work is an effort towards setting up a test bench for future development in the field of IEEE 802.15.4 wireless data transmission protocol. This work can be used for following purposes.

- Analysis of IEEE 802.15.4 protocol.
- Implementation of more efficient power saving algorithm.
- Setup can be used to compare performance of new hardware design for wireless data transmission.
- It can be used as a prototype for development of devices for commercial purposes.
- Current work can be used in other application to make data transfer wireless.

# 6. REFERENCES

[1] Institute of Electrical and Electronic Engineers, Inc., "IEEE Std. 802.15.4-2003, IEEE Standard for Information Technology – Telecommunications and Information Exchange between Systems – Local and Metropolitan Area Networks – specific Requirements – Part 15.4 : Wireless Medium Access Control (**MAC**) and Physical Layer (**PHY**) Specifications for Low Rate Wireless Personal Area Networks (**LR-WPAN**)"

[2] Rafidah Ahmed, Othman Sidek and Shukri Korakkottil Kunhi Mohd. (2009), "**Development of CRC Block on FPGA for Zigbee Standards**," IEEE Trans. Industrial Electronics, CEDEC Engineering Campus, Malaysia, 2009

[3] P Mohana, S Radha (2009), "**Realization of MAC Layer Functions of ZigBee Protocol Stack in FPGA**", International Conference on "Control Automation, Communication and energy conservation -2009", 4th-6th, June 2009

[4] Tuan Dang and Catherine Devic (2008), "**OCARI: Optimization of Communication for Ad hoc Reliable Industrial networks**", IEEE International Workshop on Performance Analysis and Enhancement of Wireless Networks, France, May 2008

[5] Panu Hamalainen, Marko Hannikainen, and Timo D. Hamalainen (2005), "**Efficient Hardware Implementation of Security Processing for IEEE 802.15.4 Wireless Networks**", Tampere University of Technology, FINLAND, 2005

[6] Rogelio N'Kenda Gaspar Neto Development of an Efficient Energy Model for the LR-WPAN, MS Thesis, University of North Carolina - Charlotte, December 2005

[7] M. Bhardwaj, T. Garnett and A. Chandrakasan, Upper bounds on the lifetime of sensor networks, IEEE International Conference on Communications, pp. 785–790, 2001

[8] D. Chen, A. K. Mok, J. Yi, M. Nixon, T. Aneweer and R. Shepard, Data collection with battery and buffer consideration in a large scale sensor network, RTCSA'05, pp. 281–284, August 2005.

[9] Y. Kiri, M. Sugano and M. Murata, Performance evaluation of intercluster multi-hop communication on large-scale sensor networks, CIT'06, pp. 215–220, September 2006

[10] J. Misic, C. J. Fung and V. B. Misic, On Node Population in a Multi-Level 802.15.4 Sensor Network, IEEE Globecom'06, November 2006

[11] J. Misic and V. B. Misic, Duty Cycle Management in Sensor Networks Based on 802.15.4 Beacon Enabled MAC, Ad Hoc and Sensor Wireless Networks Journal, pp. 207–233, 2005

[12] M. Neugebauer and K. Kabitzsch, A new protocol for a low power sensor network, IEEE International Conference on Performance, Computing, and Communications, pp. 393–399, 2004

[13] V. Rai and R. Mahapatra, Lifetime modeling of a sensor network, Proceedings of Design, Automation and Test in Europe, pp. 202–203, 2005

[14] G. F. Riley and M. Ammar, Simulating Large Networks - How Big is Big Enough?, Conference on Grand Challenges for Modeling and Sim, January 2002